# EXPERIMENTS ON RADIOACTIVITY IN A VIRTUAL LABORATORY

**Amelia Carolina Sparavigna**
Dipartimento di Fisica, Politecnico di Torino
Corso Duca degli Abruzzi 24, 10129 Torino, Italy

This paper shows how to run some experiments of physics, using a virtual laboratory. Such a laboratory is a website, equipped with objects to be measured and measuring instruments, simulated by means of Java Applets. Here we discuss in particular the case of a laboratory in which we can perform, on the Web, some experiments on radioactivity. The proposed virtual environment can be a viable alternative in the case of unavailability of a real laboratory.

**Introduction**
The numerical simulation of the behaviour of physical systems is the main activity of computational physics. Combining the computational analysis with a more specific research on computer rendering, it is possible to create many virtual environments, even fairly realistic, in which teachers and students have the opportunity to perform experiments on physics. Of course, they are not actual experiments, but are simulations of resulting phenomena that can be observed, under the change of those parameters related to the physical system under consideration.
The amount of existing simulators for physics is huge: the teacher has a wide possibility of choice among programs that can help illustrating many subjects of physics. Several of these programs are also free for use on the Web. Among the websites devoted to simulations on physics, let us note "Physics with computer" (in Spanish). The site has applets on almost all topics of physics: quite interesting are those on oscillations. Through these applets, the teacher can offer the student some practical exercises, designed to show the different topics of the theory. The website [1] has also some simple tools, virtual instruments, which simulate instruments such as scale and gauge [2].
In this paper we will show that it is possible to perform some virtual experiments within already existing pages on the Web; in this manner, the website becomes a valid replacement of a real laboratory. We will analyze in particular the case of a laboratory in which run measurements on radioactivity. Let us point out that a real laboratory, where experiences on radioactivity could be performed, must satisfy specific criteria of safety and security for researchers and students attending the lessons. To have a virtual laboratory allows running experiments on radioactivity even when the educational institution does not have room or financial resources for the equipment of the laboratory itself.

**The virtual laboratory of radioactivity**
A virtual laboratory is simply a Web page, equipped with virtual objects and tools as possible, to simulating the real ones. It can also have the same safety controls that exist in the real laboratory: when the user enters the page for instance, she/he may be forced to read the regulations concerning protection against ionizing radiations. This way of enter the site with access control is also useful as user's training. The user learns in advance the safety checks that must be carried out entering a real laboratory.
The example we show here is very simple: entering the virtual room, the student can use a Geiger counter and two radioactive sources. This is what we find at the site "Cal Poly Physics Department's Virtual Radiation Laboratory" [3] and "Applets Java" [4], where professor Peter Siegel, California State Polytechnic University, created a virtual environment for experimental work on radioactivity, radioactive decay [3] and more [4]. At [3], Siegel is telling that the virtual laboratory has the purpose to give the students "an introduction to radiation detection and data analysis without being radiated". There we find the simulation of four different detectors: a Geiger counter, two gamma detectors and a liquid scintillation detector. Websites [3-5] are displayed with the internet browsers. The Java Applet Geiger Counter was created by Andres Cardenas, from the California Polytechnic University too [5]. Let us remember that Java is an object-oriented programming language, enabling the creation of suitable applications to be implemented in many environments. When an application written in Java is installed inside a browser (Chrome, Mozilla, Explorer), it is called a Java applet.

In website [5] (Fig. 1), we can run several experiments: among them we can imagine the half-life determination of a radioactive material, for instance, or a study of the statistics of radioactive decay. The simulator can also determine a characteristic of the Geiger counters, its dead time. Let us remember that the single radioactive decay causes a discharge between the electrodes of the instrument. The discharge takes some time, during which the instrument is not able to measure any other decay: this is called the dead time of the instrument, $\tau$.

The virtual Geiger counter has two sample holders where it is possible to put two types of radioactive sources. The simulated samples are radionuclides Ba137m and Mn54 (5 $\mu$-curies) [7]. The sample holder can also be left empty, and thus the simulator measures the radioactivity of natural background. An experiment that can be performed is then the study of the natural background. Another virtual experiment that can be done with the applet [5] is the determination of the half-life of a radioactive material through the source Ba137m (half-life of about 2:55 minutes). The instrument is displaying the total elapsed time in the laboratory. Moreover, it is possible to select some specific time intervals, during which the virtual instrument evaluate the counting rate. Let us see in details some measurement that can be performed within this website [5].

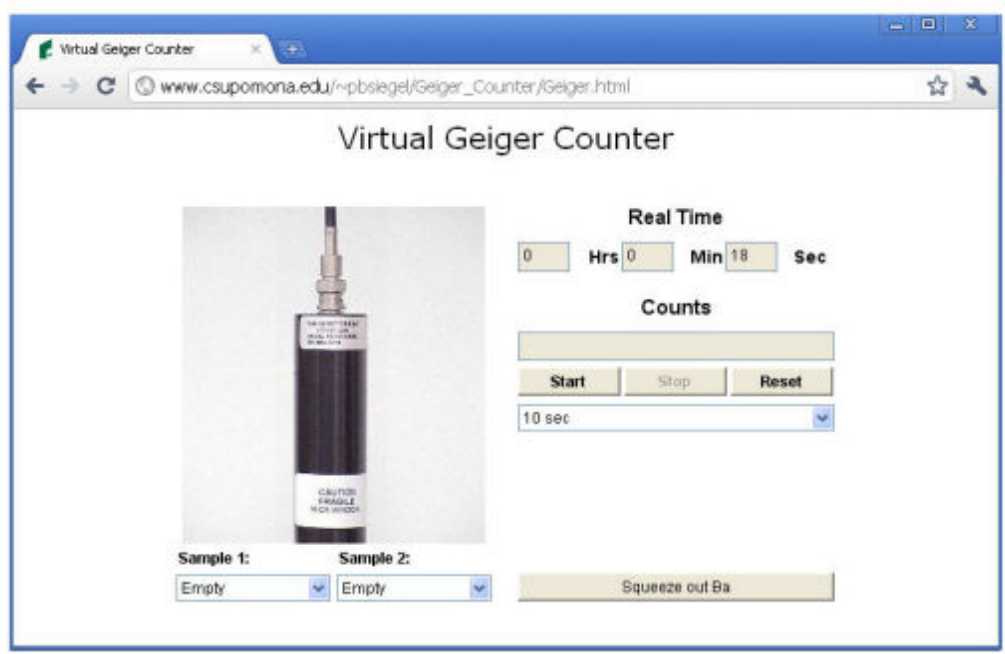

Fig.1 Screenshot of the virtual laboratory of radioactivity [5], equipped with a Geiger and two sample holders. These holders can be left empty, to measure the natural background, or loaded with two different types of radioactive sources.

**Measurement of the dead-time $\tau$.**
A Geiger-Mueller tube is a gas-filled radiation detector. It usually takes the form of a cylindrical outer shell (the cathode), with inside a sealed gas-filled space, and a central wire (the anode) held at a high voltage with respect to the cathode. The gas is generally Argon at a low pressure, with a small quantity of a quenching vapour. If a $\gamma$-ray interacts with the instrument, primarily with its walls, an energetic electron is produced that can travel through the tube. This primary electron produces ionization along its path, by collisions with the fill gas. The new electrons and the original one are accelerated, producing a cascade of ionization known as the Townsend avalanche. As a consequence, a dense sheath of ionization is created along the central wire, producing what is termed a Geiger-Mueller discharge.

The intense electric field collects electrons to the anode and repels positive ions. The electrons are collected within a few microseconds, while massive positive ions are accelerated quite slowly, due to their low mobility, towards the cathode. The ions produce a positive space charge surrounding the central wire: the temporary presence of this space charge stops additional avalanches, because it is reducing the field gradient near the anode below the avalanche threshold. As a consequence, any approaching electron does not gain enough energy to start a new avalanche. The detector is then "dead" for the time the ions migrate far enough for the inner electrode. The time required for recovery a value of the field high enough to generate a new discharge and then a new count is then the "dead time" $\tau$, which is of the order of 100 μs.

During $\tau$, the instrument is not able to measure any other decay. Let us define $n$ as the true decay rate in a fixed time interval and $R$ the counting rate obtained by the Geiger counter during the same time interval. It is assumes that $n = R/(1-R\tau)$ [6].

The classic method for measuring the dead time of the Geiger counter is the so-called "method of the two sources" [6,8,9]. Suppose that $n_1$ and $n_2$ are the true rates of the two sources and $R_1$, $R_2$ and $R_{12}$ the counting rates measured for the two sources, separately and together, respectively. We have:

$$n_1 = \frac{R_1}{1-R_1\tau} \; ; \; n_2 = \frac{R_2}{1-R_2\tau} \; ; \; n_{12} = \frac{R_{12}}{1-R_{12}\tau} \rightarrow \frac{R_{12}}{1-R_{12}\tau} = \frac{R_1}{1-R_1\tau} + \frac{R_2}{1-R_2\tau} \quad (1)$$

then::

$$\tau = \frac{R_1 R_2 - [R_1 R_2 (R_{12} - R_1)(R_{12} - R_2)]^{1/2}}{R_1 R_2 R_{12}} \quad (2)$$

The virtual laboratory [5] has two sample holders, and the user has the possibility to apply the method of the two sources, measuring the dead time. It is necessary to use the source Mn54 with a measurement time interval of 10 seconds, for instance. Loading a sample holder, leaving the other empty, the individual measured counting rates $R_1 = 2735$, $R_2 = 2737$ have been obtained. Loading both the sample holders, the counting rate $R_{12} = 5073$ is obtained. From Eq.2, $\tau = 3. \times 10^{-5}$ sec.

**Counting rate statistics**

To determine that the counting rate of radioactive decay has a Poisson distribution we use the source Mn54, in the virtual lab [5]. The run needs approximately two hours. I chose an observation period of 10 sec over which the Geiger counted the decays. I repeated the measurement of the counting rates for $N = 180$ observation periods. After having obtained 180 counting rates, I subdivided the data in class intervals. It is usually suggested to find a range that contains all the data and subdivide it into many intervals, large enough to contain several data, but small enough to show the trend of the distribution. I propose to span the range between the minimum measured value of the counting rate and the maximum value of the real rate, $n_{max} = R_{max}/(1-R_{max}\tau)$, where $R_{max}$ is the maximum value observed during the measurements. This range is then subdivided into $m$ equally spaced intervals. Since $N$ is large, we can assume $10 < m < 20$. For a given $m$, we determine the frequency histogram, according to the class intervals. The experimental frequencies can be compared with the probabilistic Poisson distribution, expressing the probability of a number of events occurring in a fixed period of time, with expected value equal to the average experimental value $\lambda$. With the 180 measurements we have the histogram in Fig.2, showing the frequency of counting rates over 15 classes, compared with the trend of the Poisson distribution (green curve).

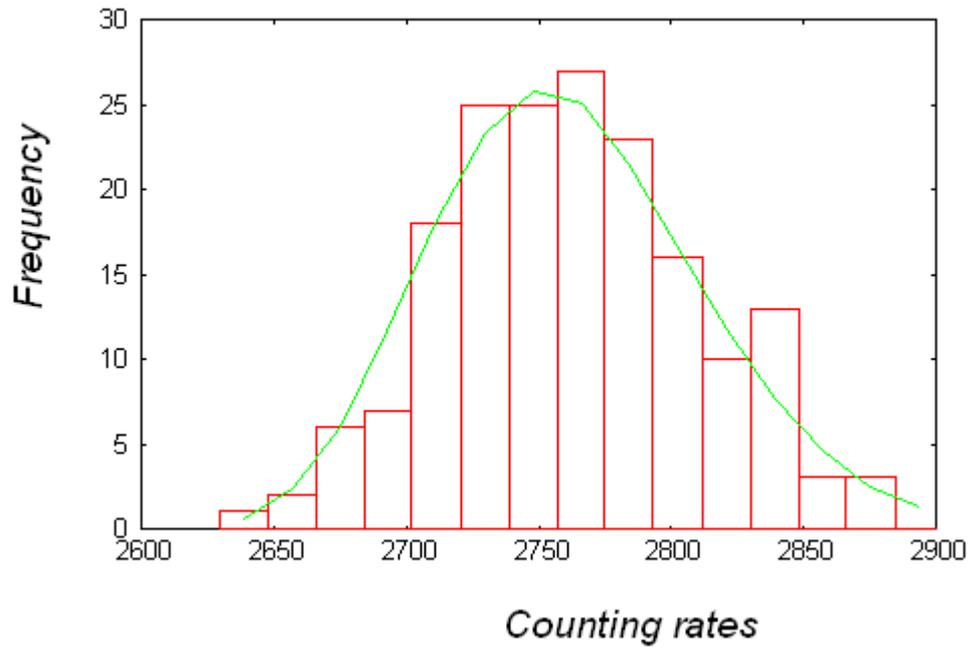

Fig.2. Measured frequencies a(red) and Poisson distribution (green).

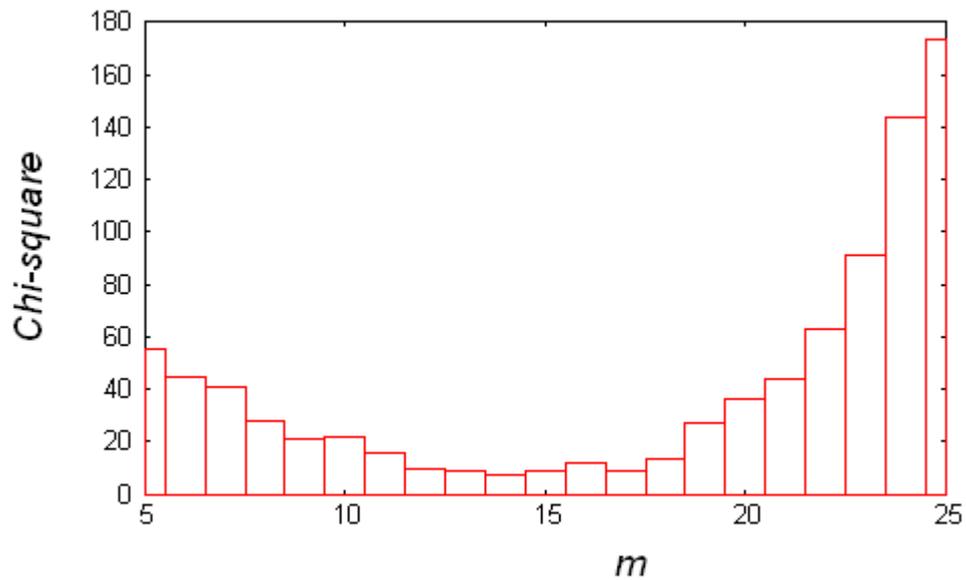

Fig.3. Value of Chi-Square (Eq.3) as a function of *m*, number of classes..

The comparison between experimental and Poisson distributions is obtained by means of the Pearson's Chi-Square test. The Pearson's test, as a test of the null hypothesis, is here applied to determine if the sample was drawn from a population with a Poisson distribution. This test is the best-known of several Chi-Squared- statistical procedures whose results are evaluated by reference to the Chi-Square distribution. For testing the counting rates, let us calculate the quantity:

$$X^2 = \sum_{i=1}^{m} \frac{(O_i - E_i)^2}{E_i}$$
(3)

where $O_i$ is the observed frequency in the *i*-th class and $E_i$ is the expected value.

The Chi-Square analysis then showed that the virtually measured counting rates obey the Poisson distribution (confidence 95%). In conclusion, the virtual run of measurements has the same behaviour as if we were in front of a Geiger counter with real radioactive sources.

Let us note that the value of $X^2$, Eq.3, depends on the number of classes $m$. We can see this dependence in Fig.3. From the figure we note that, above $m=20$, $X^2$ is rapidly growing. From $m=12$ to $m=18$, the value of $X^2$ is approximately constant. It is then natural the choice of $m=15$, that we used for the Pearson's test.

The previous discussion is proposed just as an example of data analysis: we could repeat the procedure on the same data, after correcting them from the effects of background and dead-time.

**Squeezing out Barium**

The virtual Geiger applet, which I used in the previous two experiments, works also with Barium137. As told in [4], to record the counts from Ba137m samples, it is necessary to "squeeze out" the radioactive sample. The button refreshes both sources when clicked. To imagine what this button is simulating for, we have to note that radionuclides used in radiotherapy, such as the Technetium-99m, have very short half-lives. For this reason, specific generators are used. A Technetium-99m generator, or colloquially a Technetium cow or Moly cow (for Molybdenum) is a device used to extract the metastable isotope Tc99m from a source of decaying Molybdenum-99 [10]. Mo99 has a half-life of 66 hours and can be easily moved for long distances from nuclear centers to hospitals, where its decay product, Technetium-99m, which as a half-life of only 6 hours, is extracted and used.

For physics laboratories, isotope generator kits of Barium-137m are existing [11,12].. The PASCO Cs-137/Ba-137 m isotope generator [12] is based on the original Union Carbide patented design. Each generator kit contains Cs137 and an eluting solution. The isotope parent Cs137 has a half-life of 30.1 years and decays to the metastable state of Ba137m. This further decays by gamma emission with a half-life of 2.6 min to the stable Ba137 element. As reported in [12]; during the elution, Ba137m is "milked" from the generator, leaving behind the Cs137 parent. The regeneration of the Ba137m is due to the fact that Cs137 continues to decay. In less of 1 hour, the generator is ready again. 30 minutes after elution, the activity of the Ba137m sample is reduced to a level low enough to be safe for handling.

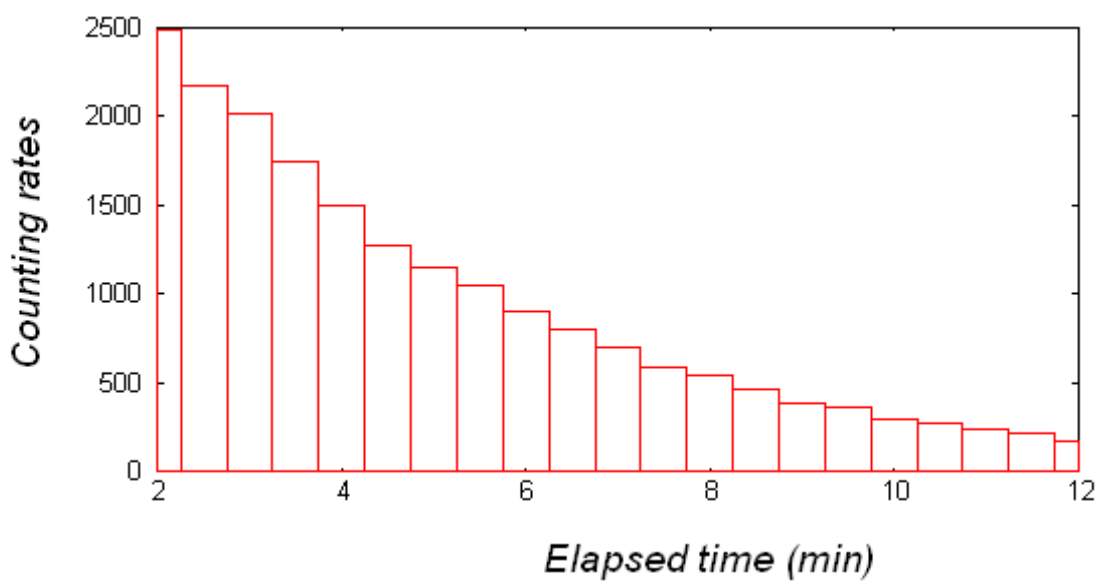

Fig.4. Counting rates of Ba137m, as obtained from [5], showing the behaviour of the decay.

After squeezing out Ba137, we can record the counting rates for 10 sec as a function of the elapsing time. Fig.4 is showing that the counting rate is decreasing with time. After two minutes of total elapsed time, I had inserted the material in one of the sample holders and measure the counting rate each 30 seconds. From the data, we can obtain the half-life.

**Conclusions**

The paper reports the description of a virtual laboratory and some of possible virtual experiences on radioactivity that we can run in it. The virtual instrument in the laboratory has been created with Java. The Java Applet in [5] is an excellent example of simulation of Geiger counter and radioactive sources. The experiments I discussed (determination of dead time, statistics of counting rates, half-time), run using the virtual instrument and sources, show that a virtual laboratory is a valid alternative to a real one.